\documentclass[preprint]{aastex}

\shorttitle{Coronagraphic Observations of $/beta$ Pic}

\shortauthors{Heap et al.}

\begin{document}

\title{STIS Coronagraphic Observations of $\beta $ Pictoris\footnote{Based on 
observations with the NASA/ESA Hubble Space Telescope, obtained at the Space 
Telescope Science Institute, which is operated by the Association of 
Universities for Research in Astronomy, Inc., under NASA contract NAS5-26555}} 

\author{Sara R. Heap, Don J. Lindler\altaffilmark{2},Thierry M. 
Lanz\altaffilmark{3}, Robert H. Cornett\altaffilmark{4}, and Ivan 
Hubeny\altaffilmark{5}} 
\affil{Laboratory for Astronomy \& Solar Physics, Code 681, NASA's
Goddard Space Flight Center, Greenbelt MD 20771}

\email{heap@stis.gsfc.nasa.gov}

\author{S.P. Maran}
\affil{Space Sciences Directorate, Code 600, NASA's
Goddard Space Flight Center, Greenbelt MD 20771 \\}
\email{maran@stis.gsfc.nasa.gov}

\and

\author{Bruce Woodgate}
\affil{Laboratory for Astronomy \& Solar Physics, Code 681, NASA's
Goddard Space Flight Center, Greenbelt MD 20771}
\email{woodgate@stis.gsfc.nasa.gov}

\altaffiltext{2}{Advanced Computer Concepts, Potomac MD 20854}
\email{lindler@stis.gsfc.nasa.gov}

\altaffiltext{3}{Department of Astronomy, University of Maryland, 
College Park MD 20742}
\email{lanz@stis.gsfc.nasa.gov}

\altaffiltext{4}{Raytheon ITSS, 4400 Forbes Blvd., Lanham MD 20706}
\email{cornett@stis.gsfc.nasa.gov}

\altaffiltext{5}{National Optical Astronomy Observatories, Tucson, AZ 85726, USA}
\email{hubeny@stis.gsfc.nasa.gov}

\date{\today }


\begin{abstract}
We present new coronagraphic images of $\beta $ Pictoris obtained with the
Space Telescope Imaging Spectrograph (STIS) in September 1997. The
high-resolution images (0\farcs1) clearly detect the circumstellar disk as
close as 0\farcs75 to the star, corresponding to a projected radius of 15
AU.  The images define the warp in the disk with
greater precision and at closer radii to $\beta$ Pic than do previous
observations. They show that the
warp can be modelled by the projection of two components: the main disk,
and a fainter component, which is inclined to the main component by
4-5$^{\circ }$ and which extends only as far as $\approx$4" from the
star. We interpret the main component as arising primarily in the outer disk
and the tilted component as defining the inner region of the disk.

The observed properties of the warped inner disk
are inconsistent with a driving force from stellar radiation. However,
warping induced by the gravitational potential of one or more planets is 
consistent with the data. Using models of planet-warped disks constructed by 
\cite{lar97}, we derive possible masses of the perturbing object.
\end{abstract}

\keywords{stars: circumstellar matter, fundamental parameters, individual
($\beta$ Pictoris), planetary systems}

\section{Introduction}

The extensive circumstellar disk of $\beta $ Pic has been subjected to
intense scrutiny and interpretation since it was discovered by \cite{smi84}.
These studies have been inspired by the possibility that the
$\beta $ Pic system might be an analogue to the solar system in its early
stages or represent a distinct branch of planetary system evolution. In
either case, $\beta $ Pic provides a laboratory in which cosmogonical
theories can be tested. Nevertheless,
fifteen years after its discovery, the $\beta $ Pic
circumstellar system remains largely a mystery. Is $\beta $ Pic a
proto-planetary environment, in which planets have not yet formed, one
where planets already exist, or a ``failed'' system, where planetesimals,
if any, cannot grow further? Spectroscopic indications of cometary bodies (e.g.
 \cite{vid94}) suggest that planet-building has progressed at
least to the planetismal stage, but resolve few issues. These questions are
thoroughly reviewed by \cite{art97} from pre-1997 data. Although there
is no direct evidence for planets around $\beta $ Pic, there are intriguing
asymmetries in the disk that may be due to gravitational perturbations by
substellar objects. The most important of these is a reported 3${{}^{\circ}}$
tilt of the inner disk ($r<50$ AU) with respect to the outer disk 
\citep{bur95,mou97}.

We present new coronagraphic observations made with the Space Telescope
Imaging Spectrograph (STIS), as installed on the Hubble Space Telescope in
February 1997. The STIS coronagrams show a greater level of detail, provide
higher image quality, and reveal the disk closer in to the star than prior
observations. These observations further characterize the tilt of the inner
disk for comparison with theory.
In \S 2, we describe the design and performance of the STIS coronagraph.
In \S 3, we discuss the $\beta $ Pic observations and data reduction
 techniques.
In \S 4, we present the observational results. The STIS images show the disk
 to be warped at close radii to
the star in  the sense that the
inner disk is tilted by 4.6 degrees with respect to the outer disk.
In \S 5, we use the size and shape of the warp to
evaluate the two proposed theories of its origin. We conclude that the
observations of the warp in the inner disk favors the existence of a planet
around $\beta $ Pic.

\section{STIS Coronagraphy}

Although designed primarily for spectroscopy, STIS easily meets the prime
requirements for stellar coronagraphy. It employs an occulting mask and Lyot
stop that work together to block out direct and scattered starlight. Because
of its high resolution, $0\farcs1=2$ AU at the Hipparcos distance of
 $\beta $ Pic (19.28 pc), it can accommodate small occulting masks, thereby allowing
examination of the $\beta $ Pic disk to within 0\farcs5, or a projected
distance of 10 AU, from the star. Most importantly, its stable point spread
function makes possible a definitive separation of the disk's light from
other components. In contrast, WPFC2 lacks a coronagraphic mode; and
ground-based coronagraphs, even with adaptive optics, are subject to
significant variations in the point-spread function ({PSF)}. Below,
we describe the design and performance of the STIS coronagraph.


The STIS design is described by Woodgate et al. (1998), its in-orbit
performance by \cite{kim98}, and its in-orbit operations by \cite{sah99} in
the {\it STIS Handbook}. Figure \ref{fig1} shows the optical layout
of the STIS coronagraph. Light entering the instrument is corrected for
telescope aberrations and for astigmatism at the STIS field point by a
two-mirror system similar to that used in COSTAR. After correction, light
passes through the coronagraphic aperture mask at the first STIS focal
plane. The beam is collimated by an off-axis ellipsoidal mirror and directed
to a flat mirror near the second pupil plane, a few millimeters beyond the
Lyot stop. The beam then goes to a folding flat and an ellipsoidal camera
that focuses the light onto the CCD detector. 


Coronagraphic observations on STIS make use of an occulting mask, called
50CORON in the {\it STIS Handbook}, which is one of 65 apertures that can be
inserted into the beam at the focal plane. Its format is clearly seen in {%
Figure \ref{fig2}, which shows a coronagraphic image of HD 60753, a V=6.61
B3III-type star having no known circumstellar structure. The square field is
51\arcsec\ across. The 50CORON aperture mask has a large rectangular bar 3%
\arcsec\ $\times $ 10\arcsec\ (top of figure),
and two tapered bars (wedges) ranging in width from 0\farcs%
5 to 3\arcsec. For these observations, the star was positioned
behind {WedgeB1.8}, a point where the bar is 1\farcs8\ wide.

The halo of the HST PSF is caused largely by light with phase errors from
the sagged outer radii of the primary mirror, with some contribution
from light diffracted from the edges of the HST primary \citep{vau91}. A
conventional Lyot stop is a baffle at a pupil plane shaped to block light
from these sources. The STIS Lyot stop, which is near but not precisely
at a pupil plane, is a circular aperture open to the central 77\% of the
beam area. It is expected to eliminate error light from both mirror sag
and the primary edge. A comparison of theoretical and observational data on
$\beta $ Pic suggests that the Lyot stop decreases the intensity of the PSF
halo by a factor of two, but further observations will be needed
to quantify the effectiveness of the Lyot stop.

The detector is a back-thinned, UV-enhanced CCD with a $1024\times 1024$
-pixel format, each pixel subtending 0\farcs0507. Hence, one CCD pixel
subtends nearly exactly 1 AU at the distance of $\beta$ Pic.
The intra-scene dynamic range of the CCD in a single exposure is 20,600;
it can be effectively increased by combining a series of exposures to raise
 the count
statistics while beating down the read noise. The low readout noise
(1.7 DN = 6.8 $e^{-}$) and dark current (6 $e^-$/pixel/hour) make it
possible to probe the $\beta $ Pic disk out to very low surface brightness.

Since the 50CORON aperture is unfiltered, the UV-enhanced CCD records light
over a very wide band, from 2,000 to 10,000 \AA . The effective {PSF} is
therefore an average of monochromatic {PSF}s, each weighted by the stellar
flux recorded by the CCD at the corresponding wavelength. The broad response
of the detector has the important advantage of blending the
diffraction rings, thereby erasing color-dependent structure in the {PSF}
and producing a smooth halo. However, the sensitivity at long wavelengths
increases the influence of scattering within the detector. 
At longer wavelengths, the wings of the {PSF} are visible out to several hundred
pixels from the stellar spectrum.


The coronagraphic image shown in Figure \ref{fig2} was obtained during the
test and evaluation of STIS following its installation on HST in February
1997.  Unfortunately, the innermost regions of the {PSF} are saturated
($>144,000e^{-}$). Nevertheless, the deep exposure is useful in revealing
structure of the {PSF} far from the star. The residual starlight outside the
occulting bar has unexpected structure. The main features are the
small ``tuft of hair'' that juts out to the upper left from the occulted
region and the two long ``stool legs'' that project downward from the
position of the occulted star. The origins of these artifacts are unknown,
but they are permanent features of the {PSF}.
Because of this general stability, the PSF can be removed or
greatly suppressed from the observations.

In principle, the {PSF} could be degraded by telescope guidance errors and
jitter and by thermal drifts within STIS. In practice, such errors are so
small that they can be ignored. However, the telescope is known to
experience focus variations (``breathing'') that take place on a timescale
of only minutes. Comparison of the three 10-min exposures that comprise the
observation of HD 60753 indicates that the effects of such variations are
insignificant compared to the intrinsic structure in the {PSF}. As a
demonstration, Figure 3 shows cross-sectional plots of the three
exposures. Only in the diffraction spikes can one
exposure be distinguished from another. We have found similar stability in
our series of exposures of $\beta $ Pic. Evidently, telescope ``breathing''
does not noticeably affect the {PSF} at distances of 1'' or more from the
star if the exposures are all taken in the same orbit.

\section{Observations and Reduction}

\subsection{Observing Strategy and Procedure}

Typically, the observing procedure for coronagraphy involves identical
observations of the target and a nearby reference star of similar magnitude
and spectral type. The intensity-scaled image of the reference star is then
subtracted from the target image in order to remove starlight in the target
image that was not already blocked by the occulting mask and Lyot stop. This
procedure was not followed in our observations of $\beta $ Pic. Instead, we
observed only the $\beta $ Pic system, but at three different spacecraft
roll angles: a roll angle that orients the $\beta $ Pic disk midway between
the telescope diffraction spikes, and at roll angles -12${{}^{\circ }}$ and
+14${{}^{\circ }}$ from the first.

This observing strategy offers two important benefits. The first is reliable
detection of structure in the $\beta $ Pic disk, if present. If an apparent
structure remains fixed in images taken at different roll angles, then the
feature must be due to the {PSF} or some detector blemish such as a ``hot''
pixel; but if the apparent structure rotates with the telescope, then it is
identified as originating in the disk. The method of roll separation may
not work close to the star if the thickness of the disk is comparable to
the linear distance moved between roll angles. In that case, any azimuthally
symmetric component of the disk would be rejected and falsely attributed to
the PSF. The method of roll separation should be checked
by direct subtraction of the PSF as defined by a suitable reference star.

The second advantage of roll separation is
suppression of noise induced by the inevitably imperfect determination of
the flat field correction. In the reduction, we flat-fielded the $\beta $
Pic images using an imaging flat of a tungsten lamp. Since $\beta $ Pic and
the tungsten lamp have different spectral flux distributions, the standard
flat will have residual errors when applied to $\beta $ Pic. These residuals
are averaged out by observing the disk at different orientations.


Another planning issue is the choice of occulting bar width to be used. The
ideal occulting bar is small enough to allow a probe of the $\beta $ Pic
disk close in to the star but large enough that light from the star does not
overwhelm the light from the disk nor saturate the detector.
We chose to make observations at two spots
along the occulting bar: at {WedgeB}1, which allows a probe to within $r=0%
\farcs5=10$ AU from the star; and at wedge B2, to within $r=1\arcsec=20$ AU.

The observations were taken on 16-17 September 1997 as part of the program
of HST early release observations (Program 7125: visits 4-6; rootname=O42V).
The observations were so scheduled because in September, the nominal roll of
the telescope orients the disk midway between the telescope diffraction
spikes. The observing program consisted of three ``visits,'' each one orbit
long, and each at a different spacecraft roll angle. In each visit, the star
was first located in the 6\arcsec\ target-acquisition ``window'' of the CCD.
Then the telescope was moved to place the star behind the {WedgeB}2
occulting bar. Next, a series of 8 five-sec exposures was recorded. Finally,
the telescope was repointed to place the star behind the {WedgeB}1 occulting
bar and two sequences of eight 3-sec exposures were taken.


\subsection{Data Reduction and Solution for the Disk}

The data were reduced with the STIS IDT version of CALSTIS \citep{lin97}.
In this program, each series of eight exposures was collected into a
data-cube, where cosmic ray hits could be identified and removed, and a
single average image formed. A bias image appropriate for a CCD gain of 4
was subtracted from the observed image. After conversion to count-rates in $%
e^{-}$/s/pixel, a dark image was subtracted. Hot pixels were identified and
removed by reference to a hot-pixel table generated from dark frames
made during the week of the observations. The resulting image was
flat-fielded with a flat generated from imaging observations of a tungsten
lamp. No sky subtraction was performed, since the sky background is very
faint. The median pixel value far from the star is 0.4 $e^{-}$/s/pixel on
the {WedgeB}2 images and 0.2 $e^{-}$/s/pixel on the {WedgeB}1 images. In any
case, any residual sky light, assumed to be uniform over the field of view,
is incorporated in the {PSF} and removed as part of the solution for the
disk.

The diffraction spikes were used to determine the precise location of the
star. Although the position of the star was stable during a given visit
(telescope roll angle), the position of the star image on the CCD array
changed by up to a pixel (0\farcs05) between visits. We therefore
geometrically registered the images at the second and third roll angles with
the first using the diffraction spikes as fiducials.

A software mask was applied to the diffraction spikes and the edges of the
occulting bar in order that they not interfere with the solution for the
brightness distribution of the disk. The effect of the software mask,
however, is to limit imaging of the disk to distances $r>0\farcs75=15$ AU ({%
WedgeB}1) and $r>1\farcs24=24$ AU ({WedgeB}2) from the star. Also in
preparation for separating the disk and stellar components of each image,
a $512\times 512$-pixel region centered on the star was extracted and
expanded to $1024\times 1024$ pixels.

Separation of the disk and star was achieved in an iterative process. As a
starting approximation, the {PSF} was assumed to be the average of the
images from each visit, each at a different orientation. The {PSF} was then
subtracted from each image and the resulting residual images aligned by
rotation and averaged to obtain a first estimate of the brightness
distribution of the disk. The disk (rotated back to the observed
orientation) was then subtracted from each image to get a new estimate of
the {PSF}. This process was repeated 100 times. The solution also involved a
rotation of the disk image by 0.94 degrees in order to align the disk along
a CCD column. Figure \ref{fig4} shows the results for the WedgeB1
observations.

No reference star was observed, so it was not possible
to make a complete check on the roll-separation procedure. Instead, we used
the image of HD 60753 (cf Fig. 2), scaled to the brightness of $\beta$ Pic,
as a template PSF. Although this star is not a good match in spectral type
to $\beta$ Pic, it produces a residual image that is very
similar to the disk images obtained by roll separation -- except in the
region close to the star, where the image of HD 60753 is saturated. Evidently,
there is no confusion problem at distances of 1\farcs5 or more from the
star.

\subsection{Evaluation of the Results}

Since there are no experimental data available to evaluate the effectiveness
of the occulting mask and Lyot stop, we compared the derived {PSF} for $%
\beta $ Pic to theoretical models. Figure \ref{fig5} compares the radial
profile of the derived {PSF} with that computed from TIM models \citep{bur93}. 
This plot demonstrates the two main advantages of coronagraphy.
First, the 1\arcsec\ occulting wedge provides a rejection factor of up to
8000. Were it not for the wedge, the star would produce count-rates of up to 
nearly a billion $e^{-}$/s/pix. But because the star is occulted, the dynamic 
range of the $\beta $ Pic scene is lowered to a point where it can easily be
accommodated by the CCD detector. For example, at $r=0\farcs5$ (10 AU),
 the occulted star contributes 26,000 $e^{-}$/s/pix, well below the full-well
capacity of the CCD (144,000 $e^{-}$ per pixel) for a 1-sec exposure. Since
the readout noise of the summed image (8 or 16 exposures for {WedgeB}2 and {%
WedgeB}1 respectively) is below 1 $e^{-}$/s/pix, its dynamic range is about $%
1\times 10^{6}$. Second, the wings of the {PSF} are a factor of two lower
than the TIM model for the telescope performance. This level of suppression
accords with the expected action of the Lyot stop, but further observations
are needed to complete the characterization of the STIS coronagraphic mode.

Figure \ref{fig6} compares the radial profiles of light from the star and
from the mid-plane of the disk. Not only does the star contribute more light to the
disk interior to $r=3$\arcsec\ but also beyond 9\arcsec ; this is because
the sky background is included in the PSF. Figure \ref{fig7}
shows the disk image with contours of the associated signal-to-noise (S/N)
ratios superposed. The errors used to compute the S/N for each pixel were
estimated from differences in the six different solutions for the disk
(observations at 3 roll angles $\times $ 2 wedge positions). As such, they
should represent a total error including both observational uncertainties
and errors in the data processing. Along the spine of the disk,
the signal-to-noise ratio exceeds 100 over the region from 30 to 150 AU
from the star. Above and below the spine of the disk, the brightness and
consequently, the S/N, drops rapidly.

\section{Observed Properties of the $\beta $ Pic Disk}

\subsection{Disk Morphology}

Figure \ref{fig8} (Plate A) shows the resulting images of the $\beta $ Pic
disk based on the WedgeB1 observations. At top is a false-color image of the disk
on a log scale. The bottom shows the disk with intensities
normalized to midplane brightness, and the vertical scale (i.e.
perpendicular to the spine of the disk) expanded by a factor of 4$\times$ in
order to show the shape of the disk more clearly. The main visual
impressions are the smoothness of the disk and the presence of a
warp close (in projection) to the star.
The smoothness of the disk in the STIS images is in
sharp contrast to previous images \citep{bur95,mou97} which
are marked by swirls and radial spikes. We interpret this
texture in previous images to incomplete elimination of the {PSF}. The
pronounced warp in the disk was detected in previous images, but only the
STIS images are able to follow it in close to the star. Below, we report on
quantitative measurements of the disk, including the radial flux gradient
and vertical flux distribution, the warp, and the innermost region of the
disk ($r<1\farcs5$) that heretofore has not been seen in images of its
dust-scattered light. To describe the disk, we use a cylindrical coordinate
system with $r$ denoting the projected distance from the star along the disk
midplane, and $z$ being the distance perpendicular to the disk plane
(``vertical'' distance).

We measured the radial profile of the disk in three different ways: ({\it i})
along the midplane defined by the outer disk, averaging over a swath $\pm
1.25$ AU high; ({\it ii}) along the curved spine (position of maximum flux),
again averaging over a swath $\pm 1.25$ AU high; and ({\it iii}) along the
midplane, but totalling the flux within a swath $\pm $100 AU high.
Figure \ref{fig9}
shows the radial brightness profile of the disk along its midplane (method
{\it i}). It shows that the brightness distribution has three segments: an
inner region with a rather flat brightness distribution, a transition
region, and an outer region where the flux falls off rapidly with increasing
radius. We fit the inner ($r=1\farcs6-3\farcs7$) and outer ($r=6\farcs7-9%
\farcs0$) segments of the profile by a power law, $I(r)=r^{\gamma }$.
Table~\ref{radind} lists the resulting values of $\gamma $ for all three
methods of measurement.

Table~\ref{comprev} compares our results with previous measurements. In
 qualitative
agreement with others, we find that: the NE and SW sides of the disk are
similar at equal radii for $r<6$\arcsec; there is a major change
in the brightness gradient at $r\approx 6$\arcsec\ = 120 AU; beyond
 $r=120$
AU, the northeast and southwest sides of the disk differ appreciably in
brightness and shape. In other respects, the STIS results are quite
 different
from those of previous studies. In the STIS images, the brightness profile
 of
the inner disk is significantly flatter -- more in accord with thermal-IR
images \citep{pan97} -- and the outer disk much steeper. Thus,
the change in slope at $r\approx 6$\arcsec\ is much more dramatic in the
STIS images ($\Delta \gamma \approx 4$). Evidently, there is a very real
change in the density distribution of dust and/or in the scattering
properties of the dust at 120 AU from the star.

Figure \ref{fig10} shows a contour plot of the disk at brightness levels
 equal to
0.1, 0.5, and 0.99 times the maximum brightness at a given projected radius.
The STIS coronagraphic images probe the disk in scattered light closer to
the star than previous observations. It is striking that the disk is seen right
up to the {WedgeB}1 (software) mask edge at a projected distance of 15 AU from 
the star. The {WedgeB}1 (line) and {WedgeB}2 (bold) observations agree quite
well except at very low surface brightness, where the disk in the {WedgeB}1
image is more distended vertically than in the {WedgeB}2 image. We interpret
this discrepancy as an indication of uncertainties in the data at low
brightness levels. The spine of the disk (99\% contours) has an approximately
constant thickness but is curved. At half-maximum brightness, the disk shows
bulges on either side (Table 3) that
are almost round and pinched off smoothly at either end. The full
width at 0.1 maximum (FW0.1M) increases with distance from the star
consistent with a constant opening angle (wedge shape).

The curvature of the spine of the disk is shown in more
detail in Figure \ref{fig11}, which shows both WedgeB1 (line) and {WedgeB}2
(bold) observations. The two observations give consistent results.
The tilt of the inner disk is seen clearly. However, the extensions of
the two sides of the inner disk do not meet at the star. Instead, the spine
of the disk dips below the disk equator close to the star ($r<1\farcs5=30$
 AU). Like \cite{kal95}, we interpret this ``wing tilt'' asymmetry as the
consequence of forward-scattering particles in a disk that is inclined to
the line of sight. Because of forward scattering, the side of the disk that is
closer to the observer will be brighter than the far side. In the case of
$\beta $ Pic, the closer side of the disk is evidently the lower (SE) side.
Midway out from the star ($r=$ 30 -- 120 AU), the spine is curved. It has
its greatest amplitude, $\Delta z\approx \pm 1.5$ AU, at $r\approx 70$ AU,
somewhat closer in than the bulges revealed by the FWHM contours in Figure
\ref{fig10}.

\subsection{Component structure of the disk}

Figure \ref{fig12} shows the vertical brightness profile at $r=90$ AU,
where the bulges are most pronounced. The asymmetrical profile
suggests the presence of two components: a main component that defines
the orientation of the disk, and a fainter component that is offset
from the main component. By assuming that each component is
symmetric vertically, we were able to separate the two components. The
results of the decomposition (Figure 13) indicate that the
fainter component is inclined by 4.6 degrees with respect to the main
component. The decomposition breaks down close to the star
($r< 50$ AU) partly because the offset of the inclined component
becomes much smaller ($<4$ CCD pixels) than the thickness of the disk.
Also, errors in the disk structure induced by roll separation of the disk
(\S 3.1) are likely to significant close to the star. Far from the star,
the solution is noisy because of the faintness of the disk.

Figure \ref{fig14} shows the radial flux distributions of the total disk
 (both main and tilted components) and tilted components. While the inclined
 component is detectable at large distances from the star, its brightness beyond
 $\approx$80 AU falls off so rapidly that it is not noticeable in the STIS images, and
the measurements are suspect. Inwards of 80 AU, its brightness declines
suggesting a central cavity especially on the NE side. At all projected
distances from the star, the southwest side of the tilted component is
the stronger, whereas the southwest side of the main component (not shown)
is somewhat weaker.

The tilted component detected by STIS shows a strong similarity to the
inner part of the disk viewed in the thermal-IR at $12 \micron$ by
\cite{pan97}.
First, its position angle is PA$_{\rm tilt}= 35.4 = 4.6+ 30.8$
degree (we use the average position angle measured by \cite{kal95} 
for the position angle of the main component) is the same as the
measured position angle of the IR disk, PA$_{\rm IR}= 35.4 \pm 0.6$ degrees
\cite{panpc}.
Second, the SW side of the tilted component is brighter than the NE side,
as is the case for the IR disk, where the SW side is nearly three times
brighter at $r\approx 3\farcs6$.
Finally, the tilted component shows evidence of a downturn in brightness
at $r<2\farcs5$, suggestive of a central cavity. Taken by itself,
this evidence would be viewed with skeptism in view of potential errors
produced in roll separation (\S 3.1). However, it is consistent with the
IR data, which is not vulnerable to such errors.

The overall, multi-wavelength structure of the disk becomes clearer
and more intriguing with the recent sub-mm maps of the $\beta$ Pictoris
system reported by \cite{hol98}. At $850 \micron$, the
disk out to $r\approx 15"$ is an elongated structure whose position angle
(PA=$32 \pm 4$ degrees) is consistent with the orientation of the main
component of the optical disk. However, there is an emission patch further
out from the star ($r=33"$) whose position angle, PA$=37 \pm 6$ degrees,
is more in line with the \textit{tilted} component seen by STIS.

Based on these comparisons with IR and sub-mm images, we identify the tilted
component as the inner part of the disk seen in the infra-red. We suggest
that past interactions in and ejection from the inner disk are responsible
for the sub-mm blob much further out. We tentatively identify the main
disk component primarily as material further out from the star than
$\approx$80 AU.

\section{Interpreting the Warp in the Disk}

There are two theories about the origin of the warping of the $\beta $%
~Pic disk. In one, the observed warp is formed by the gravitational
attraction of a planet in an orbit inclined to the dust disk \citep{bur95}.
A similar conclusion was reached by \cite{mou97}, who made
numerical simulations to show how the gravitational pull of a major planet
in orbit 3-20 AU from the star could warp the disk
further out where the bulges are seen. In the other theory developed by
\cite{arm97}, warps naturally form in the inner part of
accretion disks surrounding luminous ($\geq 10\textrm{L}_{\odot}$) pre-main
sequence stars due to a radiation-induced instability.
Both theories predict that the disk will look the same for many
thousands of years. The two theories differ, however, in the predicted
size and shape of the warp and in their assumptions about the star.
With their high image quality and high resolution, the STIS images present
stringent observational tests of the models.
Below, we compare the STIS observations to predictions of the two theories.

\subsection{Radiation-Induced Warps}

Armitage and Pringle's simulations follow the development of a radiatively
induced warp in L$\geq 10$ L$_{\odot}$ stars over the lifetime of the warp
(about 20 Myr). In its initial stages, the
warp looks like a spiral wave coming out from the star with its greatest
amplitude close to the star (see figure 3 of \cite{mal96}) The amplitude
of the warp grows to several tens of degrees in a
few Myr but then decays on a viscous timescale (a few Myr) when the disk
becomes optically thin. Over the course of time, the spiral wave washes out
so that the disk looks more symmetrical. According to this theory, the
$\beta $~Pic disk is well into the decay stage.

The inner region of the model disk, while flat and quite thin during the
 decay
phase, is tilted with respect to the outer regions of the disk, but the
tilted region only extends out to about 25 AU from the star. If the line of
nodes is aligned with the line of sight, the maximum thickness of the disk
would occur at $r=1\farcs2$. Regardless of the angle of the line of nodes,
the apparent thickness should decrease at $r>1\farcs2$. This is not the
case: the observed disk has its maximum vertical thickness in the bulge
region at $r=4\farcs5=90$ AU.

Another problem for the theory concerns the properties of the star.
In order for a radiation-induced warping mechanism to be viable, $\beta$~Pic
must be less than about 20 Myr old, and its luminosity must exceed
10 L$_{\odot}$. In interpreting the disk of $\beta $ Pic, \cite{arm97}
assumed the stellar parameters derived by \cite{lan95}.
However, these parameters have been revised because of the new, accurate
distance to $\beta $~Pic from Hipparcos \citep{cri97}, and
new, accurate spectrophotometry of $\beta $~Pic from 3200 to 7350\thinspace
 \AA\ \citep{ale96}.
The new data imply a somewhat lower stellar temperature, T$_{{\rm eff}}=7950 $
\thinspace K, a lower luminosity, $L=8$ $L_{\odot}$, than
before, and a minimum age of about 20 Myr.

We conclude that radiation-induced warping is ruled out by
observation: the observed warping extends out three times further than
predicted; at $L=8$ $L_{\odot}$, the star does not have the luminosity needed
to induce warping; and at $\geq 20$ Myr, the system is too old for warping to
be detectable.

\subsection{Warping by a Planet}

In the planetary-perturbation theory (\cite{lar97} and references therein) 
the shape of the warp depends on the orbital parameters of the planet. For example,
if the planet is in an elliptical orbit, one side of the tilted
inner disk would extend out further from the star than the other.
The radial extent of
the warp scales with $(M_{P}\,a^{2}\,t)^{2/7}$, where $M_{P}$ is the mass of
the perturber, $a$ the semimajor axis, and $t$ the age of the system. The STIS
observations showing the disk is warped out to 70\thinspace AU or more
(Mouillet et al. used 50\thinspace AU) require that:
\[
\log (M_{P}/M_{\star })+2\,\log a+\log t\approx 6.7\
\]
where $a$ is in AU, and $t$ is in years.
Furthermore, we require that $M_{P}/M_{\star}\leq 0.01$; otherwise, the perturber
would induce stellar radial-velocity variations that are not observed
($\Delta RV<1$ km s$^{-1}$ according to \cite{lagpc}).
Table 4 gives some possible values of the mass of the perturber implied by this
equation for orbital sizes ranging from 3 AU to 50 AU and for ages ranging
from 20 Myr to 100 Myr. It shows that if the $\beta$ Pic system is young
($t\approx 20$\thinspace Myr) and the companion is very close to the star
($a<$3 AU), the warp is produced by a brown dwarf or a very low-mass star.
If the perturber is further away (5--50 AU) then the
corresponding planetary mass ranges from 17.4 $M_{J}$ to 0.17 $M_{J}$, where
$M_J$ is the mass of Jupiter. If the system is more evolved ($t\approx 100$
\thinspace Myr), the same orbital parameters imply a lower planetery mass,
down to ten times the mass of the earth.

\section{Conclusions and Future Work}

1. {\it Coronagraphic performance of STIS.} Observations of $\beta $ Pic
indicate that the STIS coronagraphic aperture and Lyot stop are effective in
blocking light from the star. The rejection factor at the {PSF} core is as
high as 8000 even for occulting masks as small as 1\farcs0 in width (1\farcs5
after the software mask is applied). There is also some suppression of
the {PSF} halo, but further observations will be needed to quantify this
performance feature. The {PSF} has significant structure but is generally
stable on a timescale of minutes and days. Because of this stability,
the {PSF} can be substantially removed from the data.

2. {\it Observed properties of the }$\beta$\ {\it Pic disk}. The STIS images
clearly define the warp with
high precision and at close radii to $\beta$ Pic. They show that the
observed warping of the disk can be resolved into two disks 5$^{\circ }$ apart.
We interpret the brighter component as arising primarily
in the outer disk, and the fainter component as arising in the inner region
of the disk.

3. {\it Presence of planet(s) in the }$\beta ${\it \ Pic disk? }
The observed properties of the tilted inner disk are consistent with the
presence of a planet in the disk. Table 5 gives possible masses and
orbital sizes of the planet.  The STIS observations are consistent with the
planetary hypothesis, but before
definitive testing of planet-warped models can proceed, they first need to
be combined with thermal-emission images in the IR and sub-mm in order to
identify the major dynamical processes operating in the disk and to refine
the possible parameters of the perturber. One issue
to be resolved by the combined observations is the size of the
central clearing zone, since that gives an upper limit to the semi-major
axis of a planet's orbit. Another issue is the orbital eccentricity of the
perturbing object. An eccentric planetary orbit ($e>0.1$) has been invoked to 
explain spectral absorption features formed by ``falling evaporating bodies''. The
thermal-IR images are ideal in characterizing asymmetries in the inner disk,
which could then be used to constrain the eccentricity. We believe that
the STIS images described here, when combined IR and sub-mm images should
prove useful in constructing a comprehensive model of the disk.

\medskip

The observations presented here represent a first exploration of the $\beta$
Pic disk with STIS. Future observations should bring substantial improvement.
For example, we can check the present results by
observing a suitable reference star and
subtracting the image from the $\beta$ Pic observations to isolate the disk.
We plan to follow up the present observations with additional measurements,
including coronagraphic spectroscopy.

\acknowledgements

We thank Merle Reinhart, the program coordinator for
program 7125, for accomplishing the challenging scheduling of the $\beta $
Pic observations. We also thank Chris Burrows, Hashima Hasan, and Mark
Clampin for their work in computing the telescope point-spread function.
We gratefully acknowledge the advice of Pawel Artymowicz,
who suggested that we try decomposing the disk. We also acknowledge the
helpful criticisms and suggestions of the
anonymous referee. This study was supported by NASA via a grant to the STIS
 IDT.

\newpage
\figcaption{ Optical layout of the STIS coronagraph. \label{fig1}}

\figcaption{Format of a STIS coronagraphic image as shown by an observation
of HD 60753 (V$=6.61$, B-V$=-0.09$, spectral
type=B3III), a single star having no known circumstellar structure. The
field is 51\arcsec\ across. The data were obtained as part of SMOV program
7088 to test the coronagraphic mode of STIS. The image shown here is a raw
image, displayed on a logarithmic scale. It is one of three exposures
comprising the {WedgeB}1.8 observation of the star.\label{fig2}}

\figcaption{Stability of the {PSF} during the observation of HD 60753. Each
exposure is identified by a line type as given in the legend.
The cross-sectional profiles are taken at line
positions, $l=324$ (top) through the ``tuft of hair'' and $l=152$
 (bottom)
through the ``stool legs''. The stability
of the {PSF} is such that the three exposures cannot be distinguished
from one another. The two bright spikes in either plot are the telescope
diffraction spikes.\label{fig3}}

\figcaption{Solutions for the $\beta$ Pic disk based on WedgeB1
data. The software masks are not
applied in these displays in order to show the location of the disk relative
to the occulting
wedges and telescope diffraction spikes. Each square box is 512 pixels = 
 26"
across.\label{fig4}}

\figcaption{Performance of the STIS coronagraph as demonstrated by a
comparison of computed (dash-dot) vs. observed PSF's. The apparent
``noisiness'' of the predicted profile stems from using only 8 monochromatic
PSF's to represent the {PSF} for the clear imaging mode. The observed radial
flux profiles of the star are shown for {WedgeB}1 data (line) and {WedgeB}2
data (bold). Note the large rejection factor at the core of the {PSF} and
the modest suppression of the halo.\label{fig5}}

\figcaption{Radial flux profiles of the spine of the $\beta$ Pic disk (line)
and {PSF}(dashes). The star is at the origin. Both {WedgeB}1 (line) and
{WedgeB}2 (bold) profiles are shown.\label{fig6}}
\figcaption{Data quality of the disk image. A greyscale image shows the
WedgeB1 observations of the disk on a logarithmic stretch. Contours of
S/N =10 and 100 per pixel are overlaid.\label{fig7}}

\figcaption{STIS/CCD coronagraphic images of the $\beta$ Pic disk
(WedgeB2 observations).
The half-width of the occulted region is 0\farcs75 = 15 AU.
At top is the disk at a logarithmic
stretch. At bottom is the disk normalized to the maximum flux, with the
vertical scale expanded by 4X. \label{fig8}}

\figcaption{The midplane surface brightness of the $\beta$ Pic disk
(counts/s/pix) as
obtained from WedgeB1 observations (line) and WedgeB2 observations (bold
line). The NE extension is on top, the SW extension at bottom. The dashed
lines show power-law fits to the brightness distribution of the inner region
(1\farcs65, $r$, 3\farcs67) and outer region (6\farcs7, $r$, 9\farcs0). The
power-law indices are listed in Table~\ref{radind} and
Table~\ref{comprev}.\label{fig9}}

\figcaption{Contour plot of the $\beta$ Pic disk for both the WedgeB1
observations (line) and WedgeB2 observations (bold line). Three brightness
levels are shown: the spine of the disk ($\geq$99\% of the
maximum brightness); the full width at half maximum brightness; and
contours at 10\% of the maximum brightness.\label{fig10}}

\figcaption{Curvature of the spine of the $\beta$ Pic disk as indicated by
the {WedgeB}1 observations (line) and {WedgeB}2 observations (bold).
\label{fig11}}

\figcaption{Vertical brightness profiles of the disk at $r=90 $ AU on the
NE side (top) and SW side (bottom). The observed profile is shown in bold,
the two components as dashed lines, and the sum of the two components as
the line.
\label{fig12}}

\figcaption{Vertical offsets of the two components.
\label{fig13}}

\figcaption{Radial brightness profile of the observed (line) and tilted
(dash-dot) components.
\label{fig14}}

\clearpage

\begin{deluxetable}{lcccc}
\tablecolumns{5}
\tablecaption{ Radial Index Measurements \label{radind}}
\tablewidth{0pt}
\tablehead{
\colhead{} & \multicolumn{2}{c}{Northeast Side}
& \multicolumn{2}{c}{Southwest Side} \\
\colhead{} & \colhead{Outer} & \colhead{Inner } &
\colhead{Inner } & \colhead{Outer} \\ }\startdata
{Spine (i)} & & & & \\
\multicolumn{1}{l}{WedgeB1} & -4.59 & -1.25 & -1.13 & -5.18 \\
\multicolumn{1}{l}{WedgeB2} & -4.81 & -1.32 & -1.12 & -5.49 \\
{Maximum (ii)} & & & & \\
\multicolumn{1}{l}{WedgeB1} & -4.57 & -1.24 & -1.05 & -5.17 \\
\multicolumn{1}{l}{WedgeB2} & -4.76 & -1.30 & -1.02 & -5.44 \\
{Total (iii)} & & & & \\
\multicolumn{1}{l}{WedgeB1} & -3.83 & -0.89 & -0.96 & -4.08 \\
\multicolumn{1}{l}{WedgeB2} & -4.46 & -0.95 & -0.87 & -5.17 \\
\enddata
\tablecomments{Labels i, ii, and iii refer to methods of measurement
 described
in the text}
\end{deluxetable}

\clearpage

\begin{deluxetable}{cccc}
\tablecolumns{4}
\tablecaption{Comparison with Previous Measurements \label{comprev}}
\tablewidth{0pt}
\tablehead{
\colhead{Region (radius)} & \colhead{Northeast} & \colhead{Southwest} &
\colhead{Reference} \\ }
\startdata
1.6'' - 3.7'' & \multicolumn{1}{l}{$-1.28\pm 0.04$} & \multicolumn{1}{l}{$%
-1.12\pm 0.01$} & \multicolumn{1}{l}{This paper} \\
& & & \\
2.8'' - 6.0'' & \multicolumn{1}{l}{$-1.79\pm 0.01$} & \multicolumn{1}{l}{$%
-1.74\pm 0.04$} & \multicolumn{1}{l}{This paper} \\
3.0'' - 6.0'' & $-1.3\;\;\;$\ \ \ \ \ \ \ \ \ \ \ \ \ \ \ & $-1.3\ \ \ \
\;\;\;\;\;\ \ \ \ \ $\ \ \ \ & \multicolumn{1}{l}{MLPL 97} \\
2.8''- 6.0'' & \multicolumn{1}{l}{$-2.40\pm 0.24$} & \multicolumn{1}{l}{$%
-2.47\pm 0.36$} & \multicolumn{1}{l}{KJ 95} \\
2.4''- 5.5'' & \multicolumn{1}{l}{$-2.38\pm 0.72$} & \multicolumn{1}{l}{$%
-1.91\pm 0.89$} & \multicolumn{1}{l}{GDC 93} \\
& & & \\
6.7'' - 9.0'' & \multicolumn{1}{l}{$-4.80\pm 0.1$} & \multicolumn{1}{l}{$%
-5.50\pm 0.1$} & \multicolumn{1}{l}{This paper} \\
$>$6'' & $-3.5$ \ \ \ \ \ \ \ \ \ \ \ \ \ & $-3.5$ \ \ \ \ \ \ \ \ \ \ \
\ \ & \multicolumn{1}{l}{MLPL 97} \\
6.0''- 16.0'' & \multicolumn{1}{l}{$-3.76\pm 0.05$} & \multicolumn{1}{l}{$%
-4.07\pm 0.05$} & \multicolumn{1}{l}{KJ 95} \\
6.0''- 16.0'' & \multicolumn{1}{l}{$-3.508\pm 0.0003$} &
\multicolumn{1}{l}{$-4.182\pm 0.0004$} & \multicolumn{1}{l}{GDC 93} \\
\enddata

\tablecomments{MLPL 97 = \cite{mou97}, KJ 95 = \cite{kal95},
GDC 93 = \cite{gol93}}
\end{deluxetable}

\clearpage

\begin{deluxetable}{ccc}
\tablecolumns{3}
\tablecaption{Thickness of Disk (in AU) \label{vertext}}
\tablewidth{0pt}
\tablehead{
\colhead{R (AU)} & \colhead{FWHM} & \colhead{FW0.1M} \\}
\startdata
20 & 17 & 34 \\
50 & 16 & 45 \\
100 & 18 & 54 \\
120 & 15 & 55 \\
200 & 22 & 66 \\

\enddata
\end{deluxetable}

\clearpage
%
\begin{deluxetable}{cccc}
\tablecolumns{4}
\tablecaption{Possible Mass of Perturbing Object (in M$_J$) \label{posspar}}
\tablewidth{0pt}
\tablehead{
\colhead{} & \multicolumn{3}{c}{Age (Myr)} \\
\colhead{a (AU)} & \colhead{20}& \colhead{50}& \colhead{100} \\}
\startdata
3 & 48 & 19.4 & 9.7 \\
5 & 17.4 & 7.0 & 3.5 \\
10 & 4.4 & 1.7 & 0.87 \\
15 & 1.9 & 0.77 & 0.39 \\
20 & 1.09 & 0.44 & 0.22 \\
30 & 0.48 & 0.19 & 0.10 \\
50 & 0.17 & 0.07 & 0.035 \\
\enddata

\end{deluxetable}

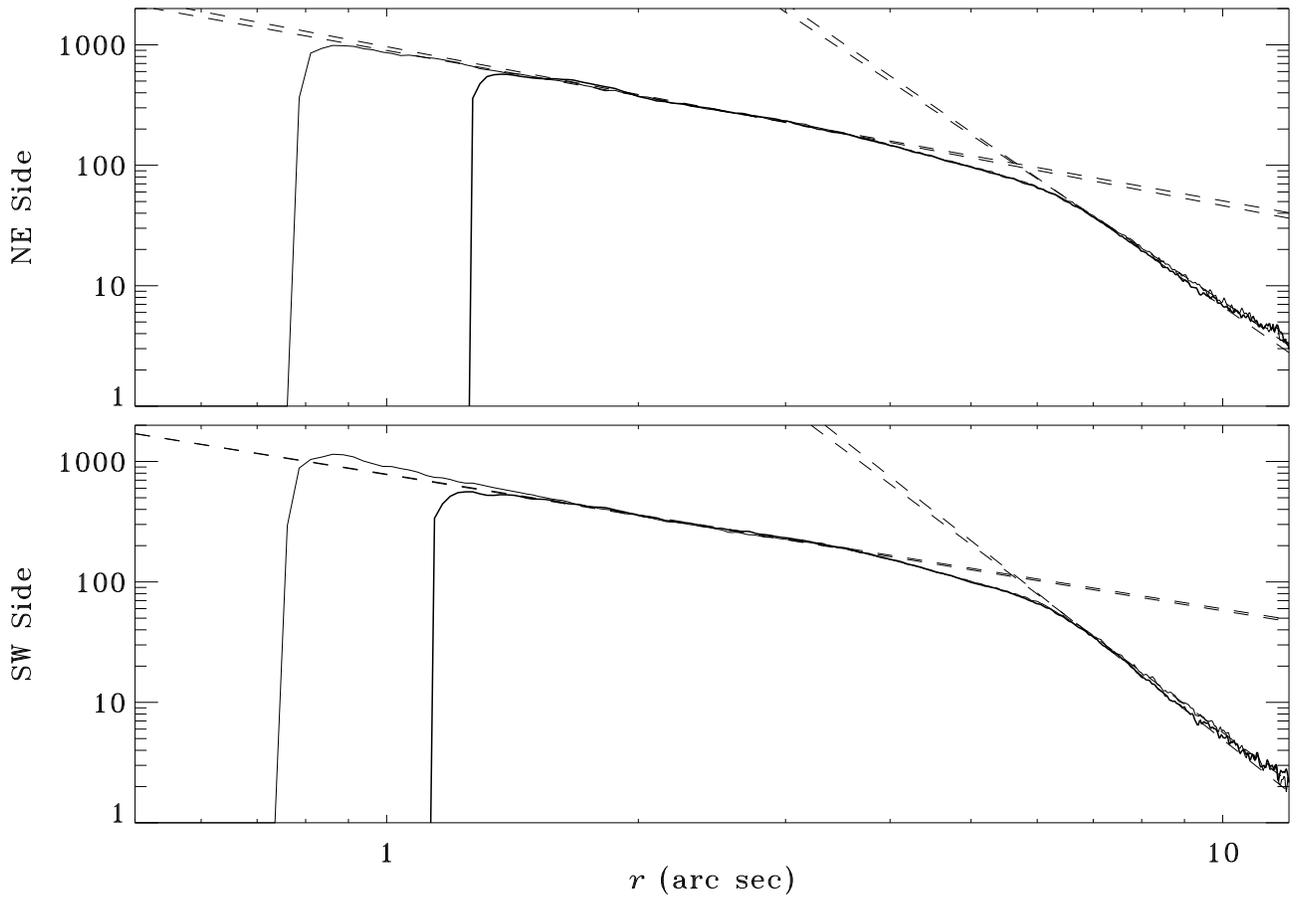
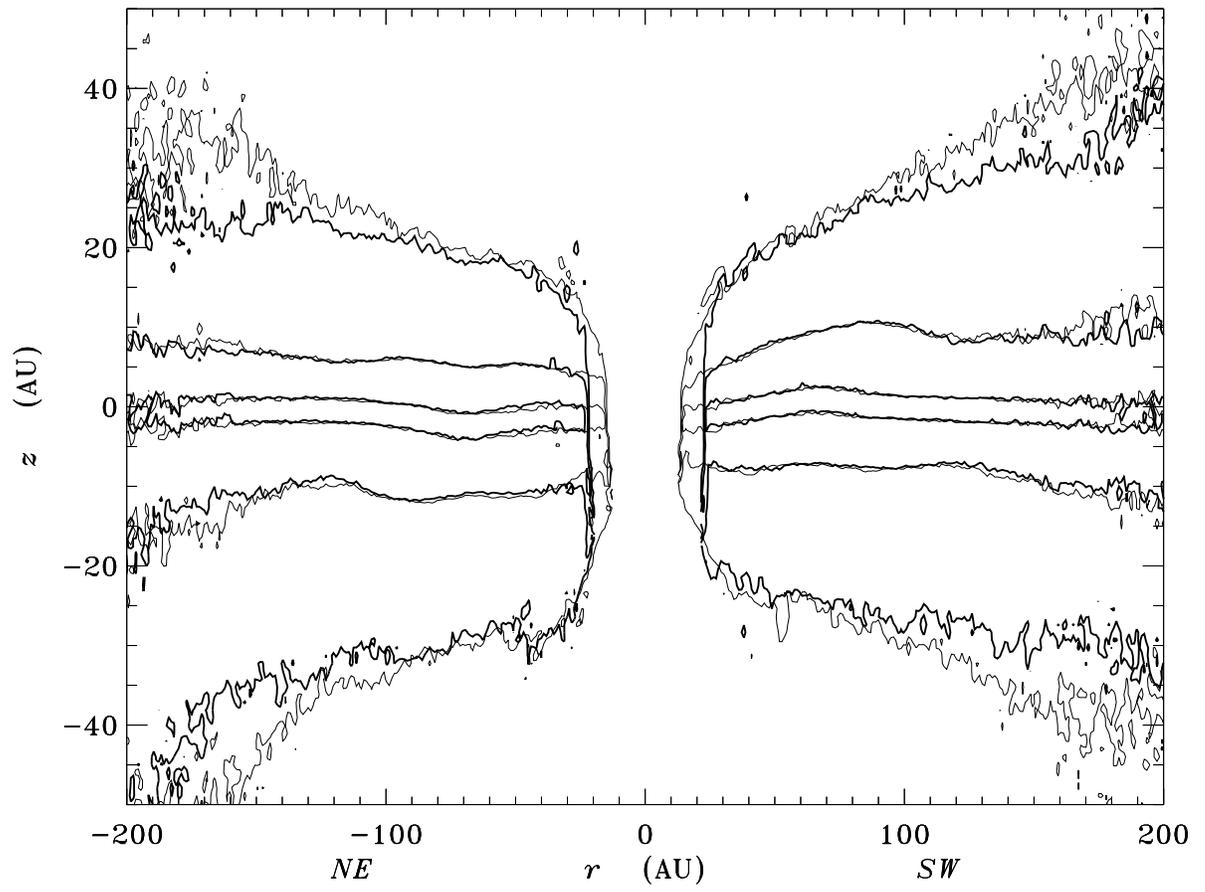
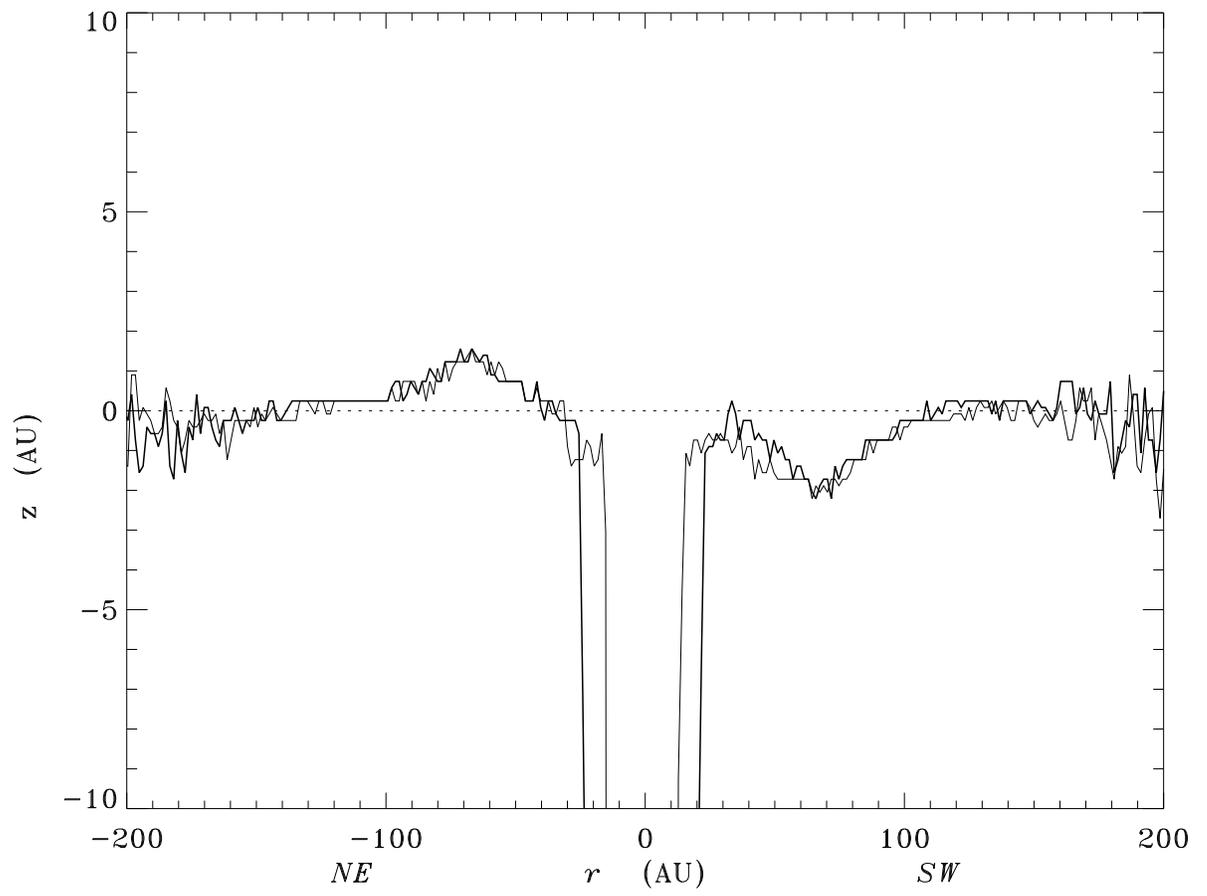
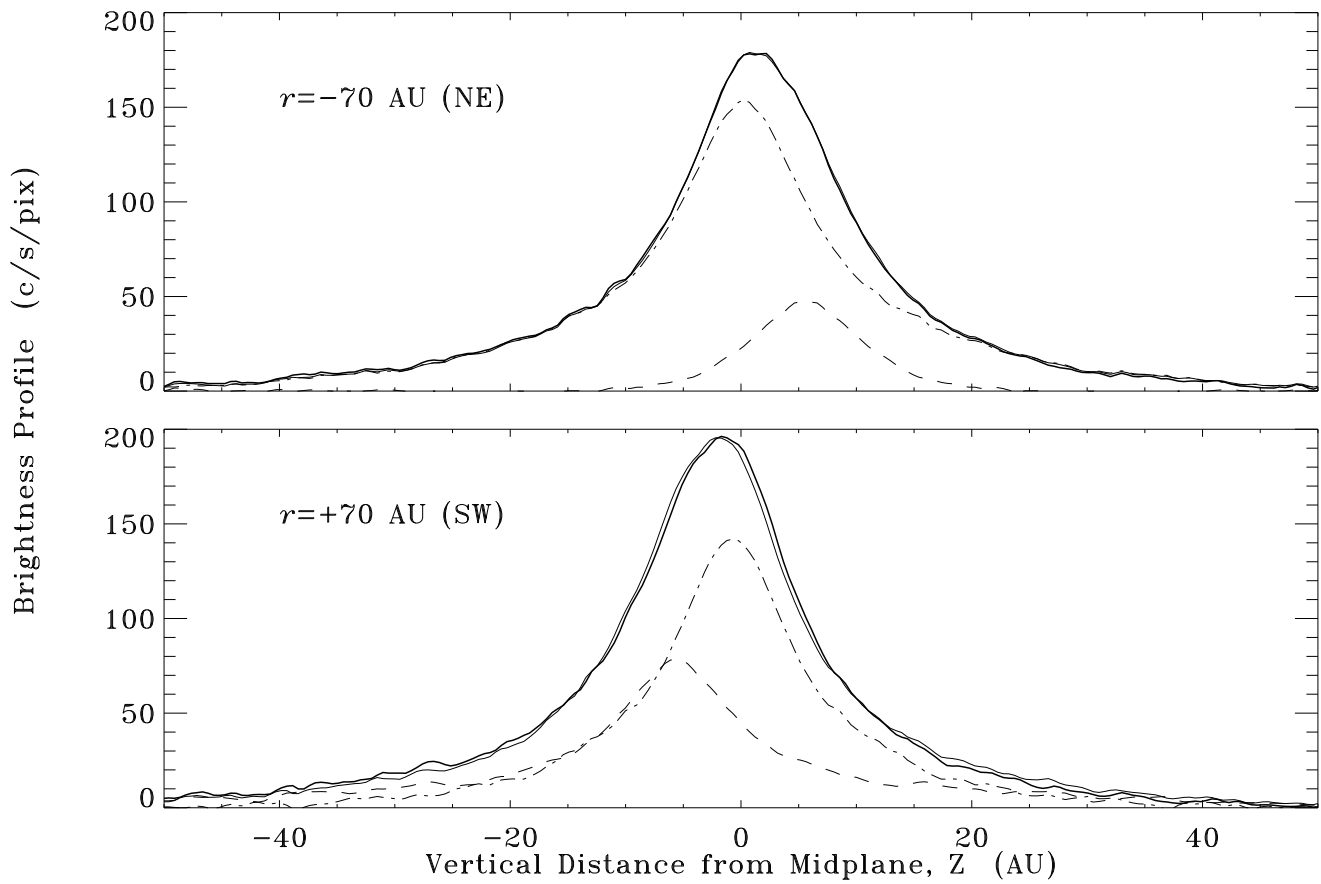
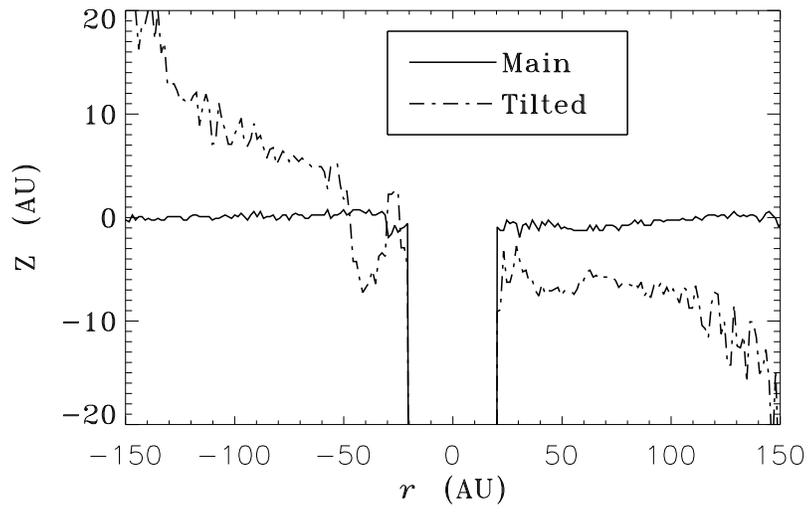
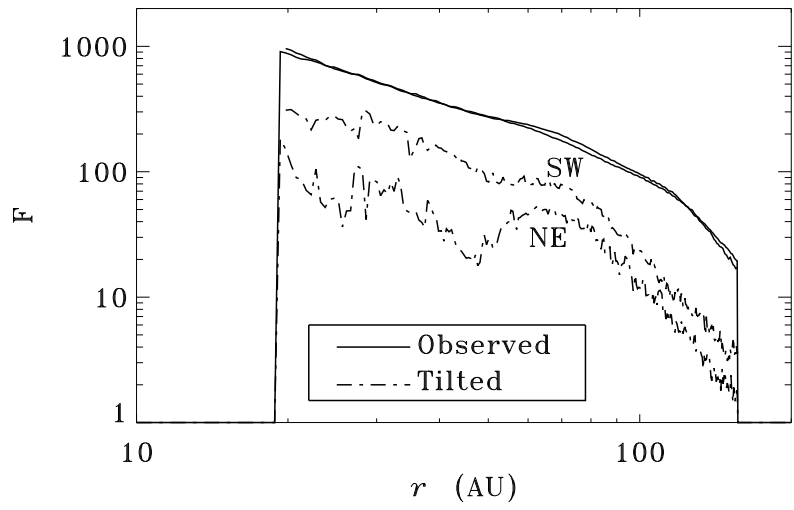
\end{document}